\documentclass{aa}
\def\msol{{M}_{\odot}}

\def\grsim{\,\lower 1mm \hbox{\ueber{\sim}{>}}\,}
\def\lesssim{\,\lower 1mm \hbox{\ueber{\sim}{<}}\,}
\input psfig
\begin{document}

   \thesaurus{03     % A&A Section 3: extragalactic
              ( 11.03.4; %{\bf Galaxies: clusters: individual:} $\ldots$
                11.09.3; %intergalactic medium
                12.03.3; %Cosmology: observations
               12.04.1; %dark matter
%                11.01.2; %Galaxies: active
                13.25.2)} % X-rays: galaxies
%               13.25.3)} % X-rays: general

\title{X-ray and optical observations of three clusters of galaxies:
Abell 901, Abell 1437, and Abell 3570\thanks{Partly
based on observations collected at the European Southern Observatory}} 

   %%  \subtitle{Test}

   \author{
                Sabine Schindler 
	}

%   \offprints{Sabine Schindler}
   \institute{
               Astrophysics Research Institute, 
               Liverpool John Moores University, 
	       Twelve Quays House,
               Egerton Wharf,
               Birkenhead CH41 1LD
               United Kingdom;
               e-mail: {\tt sas@staru1.livjm.ac.uk}
              }
   \date{}
%\authorrunning {Sabine Schindler et al.}
\titlerunning {X-ray and 
optical observations of three clusters of galaxies}

   \maketitle

   \begin{abstract}
We analyse three clusters of galaxies, Abell 901 (z=0.17), Abell 1437
(z=0.13) and Abell 3570 (z=0.037).
They have low to intermediate X-ray fluxes and  
an irregular morphology in the ROSAT All-Sky Survey (RASS). 
These clusters are chosen to
test the abilities and limitations of the RASS in terms of cluster
fluxes and cluster morphologies. Therefore some ``worst'' cases are
used here. X-ray observations with the 
ROSAT/HRI and optical spectroscopic observations are carried out.
The ROSAT/HRI observations, which have a much better spatial
resolution than the RASS, reveal in two of the three cases a
significantly 
different morphology than seen in the RASS. The reasons are point
sources which could not be resolved in the RASS and were therefore
confused with the cluster emission. For A3570 we could confirm the
relaxed state of the cluster 
by the optical determination of a small velocity dispersion.
In the cluster with the lowest flux (Abell 901) the
countrate measurement is strongly affected by point sources, in the two
other cases the countrate measurements of the RASS are reliable,
i.e. they are reproduced by the ROSAT/HRI measurement.
We conclude that for clusters with a flux of a few times $10^{-12}$
erg/cm$^2$/s or smaller, which show at the same time a non-relaxed morphology,
the flux measurement of the RASS can be seriously affected by
fore- or background sources. We point out that an all-sky survey of 
a second ABRIXAS mission
would provide a much clearer source distinction for low-flux clusters
and thus a much improved countrate determination.

   \end{abstract}

%
%________________________________________________________________
%

\section{Introduction}

Clusters of galaxies are excellent probes for cosmological theories.
Different
cosmological models predict vastly different properties for the
clusters as a function of redshift.   
Therefore relations of cluster
properties, their distributions and evolution 
can be used to constrain cosmological parameters. 
Selection according to X-ray luminosity is a good way to find the
highest mass concentrations because of a relatively well defined
correlation between the X-ray luminosity and the total cluster mass
(Reiprich \& B\"ohringer 1999; Schindler 1999). 
A second advantage of high luminosity clusters is, of course, 
that they require
shorter exposure times than their fainter counterparts to get the same numbers
of photons. Therefore many X-ray studies have concentrated so far on
luminous clusters.

But also low-luminosity clusters 
are interesting for different
reasons. For example, low-luminosity clusters
are the links between massive clusters and
groups of galaxies. Groups differ in various properties from clusters. Their 
$L_X-T$ relation is steeper than the one for clusters (compare Ponman et
al. 1996 with Arnaud \& Evrard 1999), 
their gas mass fraction is smaller (compare Pildis et al. 1995 and
Ettori \& Fabian 1999 or Mohr et al. 1999) and
their silicon abundance is lower (Fukazawa et al. 1998).
To investigate the physics causing these differences it is important
to understand how these properties change with mass or X-ray luminosity.

Low-luminosity clusters are also needed to determine the slope of
various relations well. 
With a long leverage the slope of a relation can be
determined more accurately, therefore the luminosities should
preferably span
several orders of magnitude.
For other studies, e.g. the determination of the correlation function
or distribution of properties, large samples of clusters are desirable
to get the most accurate results. Therefore, when large numbers
are required also faint clusters must be taken into account. 
%As the luminosity function is quite steep (e.g. De Grandi
%et al. 1999) faint clusters are much more numerous than luminous
%ones. 

\begin{table*}[htbp]
\begin{center}
\begin{tabular}{|c|c|c|c|c|}
\hline
cluster & \multispan 2 ROSAT/HRI \vrule
        & \multispan 2 {ESO3.6m/MEFOS} \vrule \\
        & exposure time & date &exposure time& date\\
\hline
A901    & 12680 s  & 15/05/97 - 01/06/97 &9000 s& 15/05/96\\
A1437   & 16210 s  & 30/06/97 - 02/07/97 &-    & -        \\
A3570   & 18860 s  & 04/02/98 - 05/02/98 &5400 s& 15/05/96\\
\hline
\end{tabular}
\end{center}
\caption{X-ray and optical observations of the three clusters}
\label{tab:obs}
\end{table*}

Nearby low-luminosity clusters sometimes provide problems. Not only is
the countrate determination
more sensitive to the background subtraction, but also contamination
by other fore- and background sources can become critical, in
particular when the cluster shows
asymmetries and substructure. 
The degree of source confusion affecting morphological studies and
countrate determinations depends on the sensitivity and the 
resolution of the detector.
Already the factor of 5 between the point spread function (PSF) 
of the ROSAT/PSPC
and the ROSAT/HRI can make a big difference, in particular for low-flux
clusters (compare the ROSAT/PSPC observation of Cl0939+4713 (Schindler
\& Wambsganss 1996) and the corresponding  ROSAT/HRI observation
(Schindler et al. 1998)). In the RASS (Voges et al. 1996)  the spatial
resolution is even more critical 
because the source is not always in the centre of the field-of-view,
where the detector has the best spatial resolution, like in
most pointed observations. But the source  
is observed at different off-axis angles as the detector scans
the sky. 
The radius including 50\% of the photons increases by about a factor
of 13 when moving from the centre to the border of the ROSAT/PSPC.
As the final image of a 
source is composed of many PSFs the final 50\% radius in the RASS is
about 80$\arcsec$, i.e.
about six times larger than the on-axis PSPC resolution (depending
also on the energy).

Clusters which have low fluxes because they are far away, 
are also very interesting, in particular for cosmological applications. 
But their extent in the sky is much smaller
than for nearby clusters and thus they are less affected by
source confusion. Therefore, we concentrate this investigation on
nearby clusters with low to intermediate luminosity.

Three clusters -- A901, A1437 and A3570
-- were selected for a 
detailed look with the high resolution of the ROSAT/HRI and for
optical observations. Each of the
clusters shows irregular structure in the RASS. This investigation
does of course not show
representative RASS clusters, but it is a ``worst-case'' study.
For RASS clusters with high flux and/or regular shape countrate determination
and morphological analyses are straightforward and reliable.

The paper is organised as follows. 
After a description of the observations (Sect. 2) we present the
analysis of the optical data in Sect. 3 and the analysis of the X-ray
data in Sect. 4. Finally, Sect. 5 gives our summary and conclusions.
Throughout this paper we use $\rm{H}_0 = 50$ km/s/Mpc and $\rm{q}_0=0.5$.

\section{Observations}

Three clusters were selected which have low to intermediate X-ray
luminosity and show at the same time a very irregular structure
in the RASS: A901, A1437, and A3570 (the RASS images will be published
by B\"ohringer et al.). 
For this analysis X-ray and optical observations were conducted.
The X-ray observations were carried out with the ROSAT/HRI (Tr\"umper
1983) and analysed with EXSAS. 
%They enable us to do a morphological
%analysis and hence determine the dynamical state.
The optical observations consist of
spectra taken with MEFOS (Felenbok et al. 1997) at the ESO3.6m telescope.
The analysis of the spectra was carried out with Figaro. The spectra
determine the distance of the cluster in the case of A901 and for
A3570 a velocity distribution.
%, which gives another hint on the
%dynamical state.
Details of the optical and X-ray observations
are listed in Table~\ref{tab:obs}.

\section{Optical analysis}

For two of the clusters -- A901 and A3570 -- optical spectra could be
taken. In A901 redshifts of three galaxies could be well
determined. These three galaxies are  at similar redshifts (see
Table~\ref{tab:redshift}). We derive a cluster
redshift of $z_{A901}=0.17$. Unfortunately, the number of galaxies is
not sufficient to estimate the velocity distribution.

\begin{figure} 
\psfig{figure=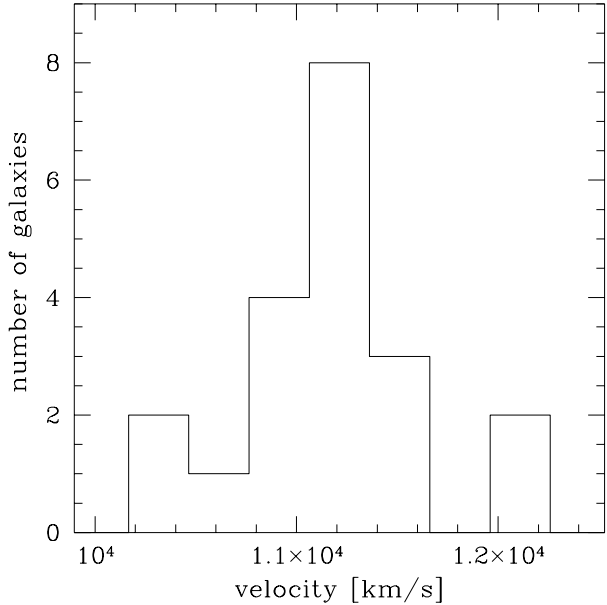,width=8cm,clip=}
\caption[]{Velocity distribution of 20 galaxies in A3570
}
\label{fig:veldisp}
\end{figure}

In A3570 we found 17 cluster member galaxies (see
Table~\ref{tab:redshift}) in addition to the three
galaxies known previously 
(Postman \& Lauer 1995). From these 20 galaxies we
derive a cluster redshift $z_{A3570}=0.0375$
(very similar to the redshift 0.0372 by
Abell et al. 1989). The velocity distribution is shown in
Fig~\ref{fig:veldisp}. 
The velocity dispersion based on these 20 galaxies 
$\sigma=460$km/s is relatively small. Such a small velocity
dispersion is a first hint for a virialised state of the cluster. If
merging of subclusters would take place and not all of the mergers would
happen perpendicular to the line-of-sight a broad velocity distribution
would be expected. 
Assuming virial equilibrium we estimate the cluster mass from the
velocity dispersion and the estimate of the virial radius by Girardi
et al. (1998) to $M_{opt}(A3570)=2.7\times10^{14}\msol$.

\begin{table}[htbp]
\begin{center}
\begin{tabular}{|c|c|c|c|c|}
\hline
cluster & $\alpha$(J2000) & $\delta$(J2000) & redshift \\
\hline
A901 
&09 54 57.5 &  -10 11 50 &0.171 \\ %25
&09 55 38.2 &  -10 10 19 &0.169 \\ %26
&09 56 53.2 &  -10 10 14 &0.164 \\ %6
\hline
A3570
&13 45 17.2 &  -38 10 21 &0.0372 \\ %26
&13 45 46.0 &  -37 56 45 &0.0406 \\ %24
&13 45 49.7 &  -37 37 18 &0.0365 \\ %20
&13 46 09.9 &  -38 00 01 &0.0364 \\ %25
&13 46 18.5 &  -38 20 18 &0.0405 \\ %29
&13 46 54.0 &  -37 51 48 &0.0376 \\ %19
&13 47 04.1 &  -38 03 57 &0.0348 \\ %28
&13 47 08.9 &  -37 55 23 &0.0366 \\ %27
&13 47 11.7 &  -38 07 33 &0.0379 \\ %30
&13 47 23.9 &  -37 51 00 &0.0390 \\ %18
&13 47 27.3 &  -37 57 54 &0.0390 \\ %2
&13 47 45.8 &  -37 50 41 &0.0376 \\ %10
&13 47 51.0 &  -38 00 52 &0.0383 \\ %5
&13 47 57.5 &  -37 56 29 &0.0373 \\ %6
&13 48 07.2 &  -38 16 47 &0.0346 \\ %3
&13 48 38.2 &  -37 43 41 &0.0355 \\ %12
&13 49 12.0 &  -37 55 54 &0.0370 \\ %8
\hline
\end{tabular}
\end{center}
\caption{Redshifts and positions of galaxies in the clusters A901 and A3570.}
\label{tab:redshift}
\end{table}

\section{X-ray analysis}

\subsection{Abell 901}

%A901 is a cluster of galaxies of richness class 1.

\begin{figure*} 
\begin{tabular}{cc}   
\psfig{figure=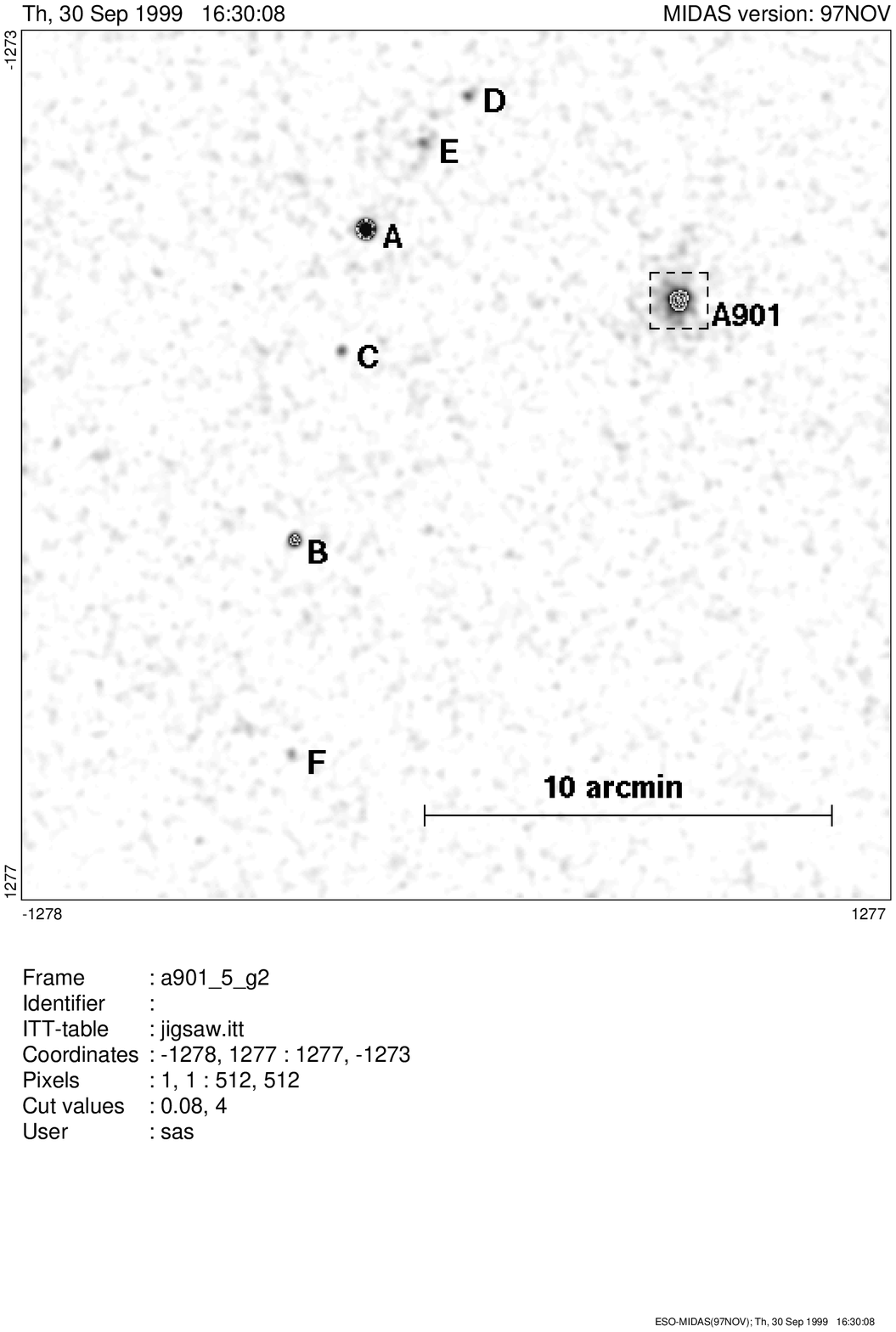,width=8.3cm,clip=} \put(-210.,200.){a} &
\psfig{figure=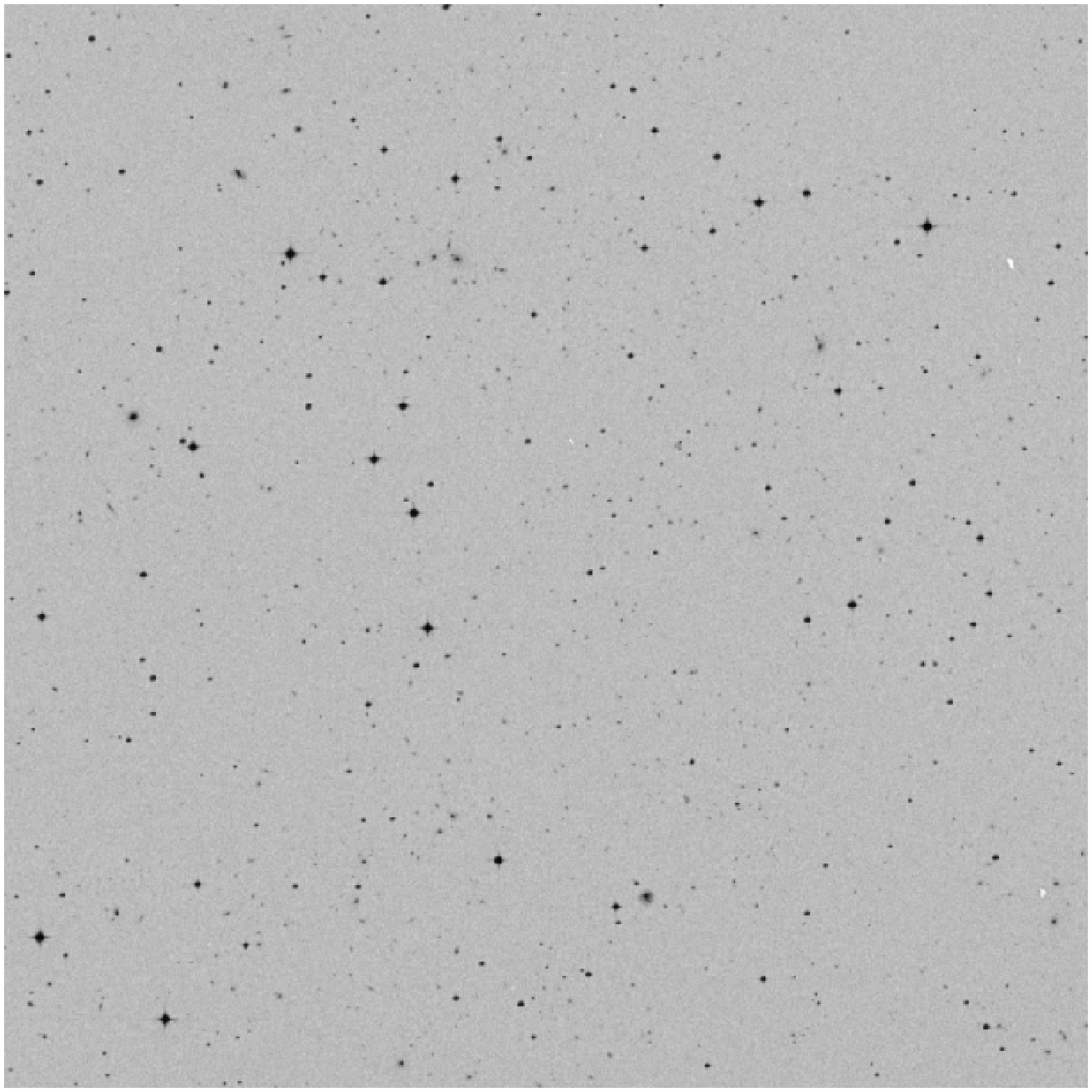,width=8.cm,clip=} \put(-210.,200.){b} 
\put(-54.,146.){A901}
\put(-130.,165.){A}
\put(-141.,137.){C}
\put(-119.,195.){E}
\cr
\end{tabular}
\caption[]{Region around A901. a) X-ray 
image taken with the ROSAT/HRI. Apart from the extended emission of
A901 six point-like sources (A-F) are visible (see also
Table~\ref{tab:point}). 
b) Optical image from
the Digitized Sky Survey of the same size as a). 
The X-ray sources A, C and F could be identified
with optical counterparts (see Table~\ref{tab:point}).
}
\label{fig:a901_over}
\end{figure*}

A ROSAT/HRI image of a  21$\arcmin$ region around A901 shows
several sources (Fig.~\ref{fig:a901_over}). A901 turns out to
have a very compact structure contrary to the
previous conclusion from the RASS. Ebeling et al. (1996) list A901
as a double cluster. Their
so-called ``brighter subcluster'' turns out to be a number of
point-like sources, while the true cluster emission is
what they call the ``fainter subcluster''. 
In Fig.~\ref{fig:a901_over}a six X-ray
point sources with a signal-to-noise ratio of at least 3 
are indicated. For three of them optical
counterparts can be found on ROE/NRL COSMOS finding charts 
(see  Fig.~\ref{fig:a901_over}b and
Table~\ref{tab:point}). Unfortunately, for none of these counterparts 
optical spectra could 
be taken because the optical observations were carried
out before the X-ray observations. At the position of A902
($2\arcmin$ West of F) no X-ray emission can be detected.

\begin{figure*} 
\begin{tabular}{cc}   
\psfig{figure=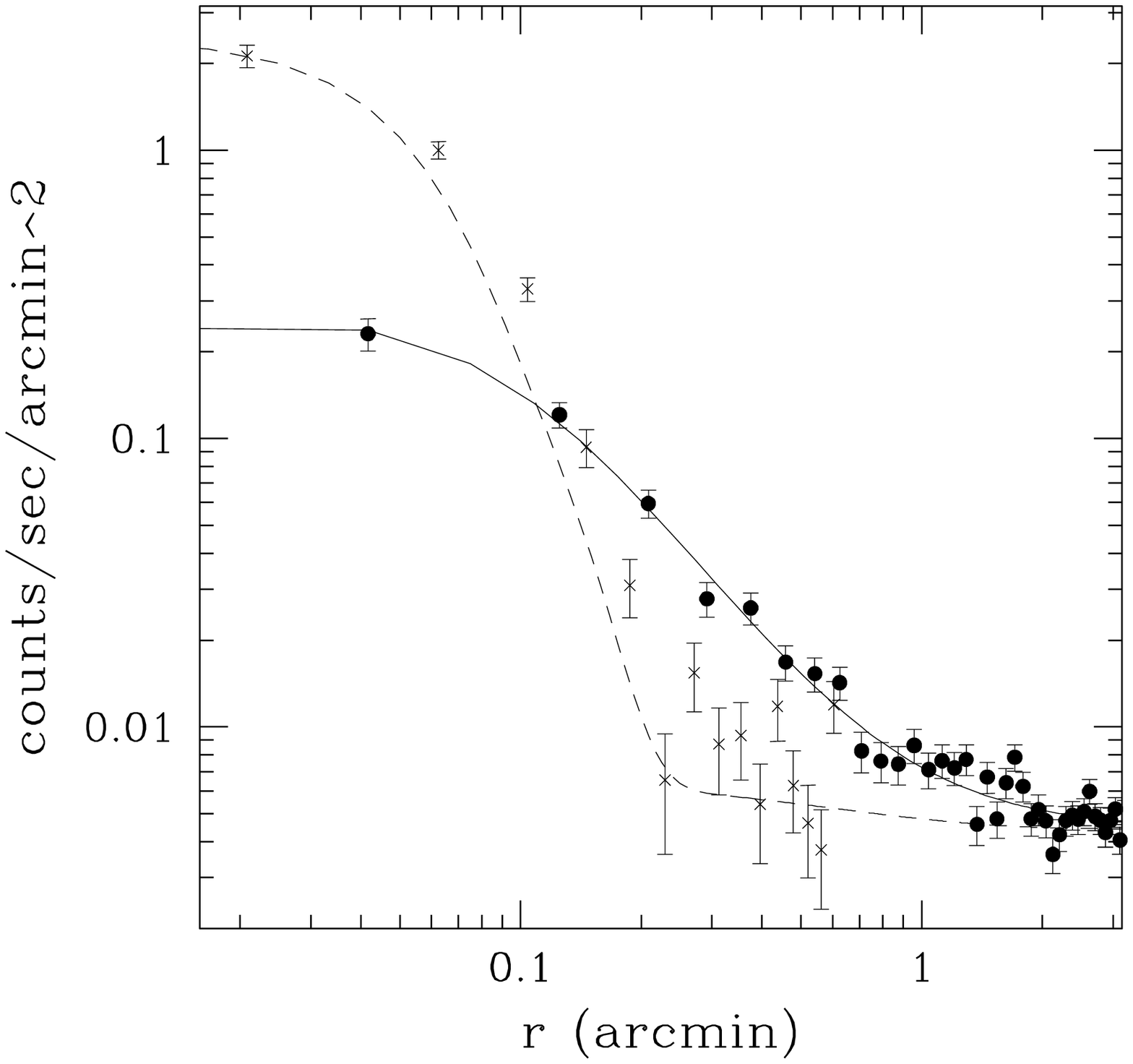,width=8.4cm,clip=} \put(-30.,190.){A901} &
\psfig{figure=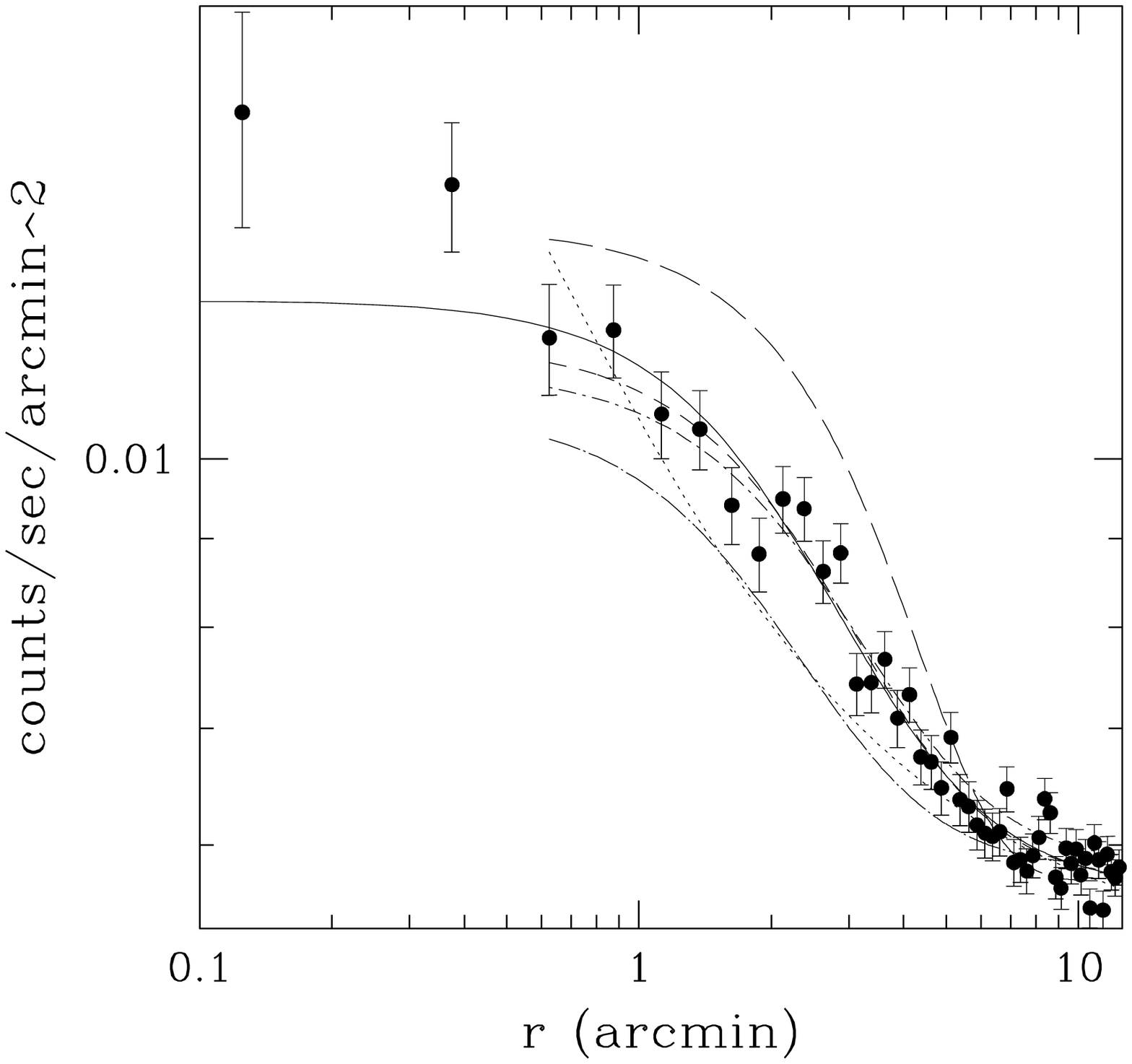,width=8.4cm,clip=} \put(-40.,190.){A1437} 
\cr
\end{tabular}
\caption[]{
Radial profiles of two clusters. Left: A901. The filled circles show the
cluster profile with the corresponding fit (full line). The crosses
show the profile of the source A. For comparison the on-axis PSF 
of the ROSAT/HRI normalised to the central bin is shown
(dashed line). As the source 
A is not exactly in the centre of the pointing but 6$\arcmin$ away,
the profile is expected to be slightly more extended than the dashed
line as seen here. Therefore A is probably a point source.
Right: A1437. Due to the
asymmetry of the cluster, fit curves for
different $90^\circ$sectors are derived, 
sector $ 10^\circ - 100^\circ$ (dash-dotted line) (N over E),
sector $100^\circ - 190^\circ$ (dotted line),
sector $190^\circ - 280^\circ$ (long-dashed line),
sector $280^\circ - 370^\circ$ (long dash-dotted line).
For clarity only the data points of the profile averaged over all
sectors are shown. The corresponding fit curves are the full line for
a fit of all data points and the short-dashed line for a fit excluding the
two innermost data points. To make these different inner radii
visible in the figure, the curves start at the radius that was used as
the inner radius for the fit.
}
\label{fig:prof}
\end{figure*}

The optical centre of A901 determined by Abell et al. (1989) 
is located between the X-ray position of A901 and source A. 
%he reason for the Abell position is a
%oncentration of galaxies around A. 
To test the extent of 
the X-ray emission of A we derive an
X-ray profile (see Fig.~\ref{fig:prof}a). The profile is compared with
the on-axis PSF of the ROSAT/HRI. The profile of A
is only slightly more extended than the on-axis PSF, 
which is expected for a point source 6$\arcmin$ away from the
pointing position. Therefore, we
conclude that most likely the X-ray emission from A is point-like and
therefore not (sub-)cluster emission, but probably emission from an active
nucleus in the centre of a galaxy. The  most likely candidate for this
AGN is a galaxy of 16.6$^m$ in B (see Table~\ref{tab:point}). 
This galaxy is located at a distance
of 5$\arcsec$ from the cluster emission -- a distance smaller than
the pointing accuracy of ROSAT. 

\begin{figure} 
\psfig{figure=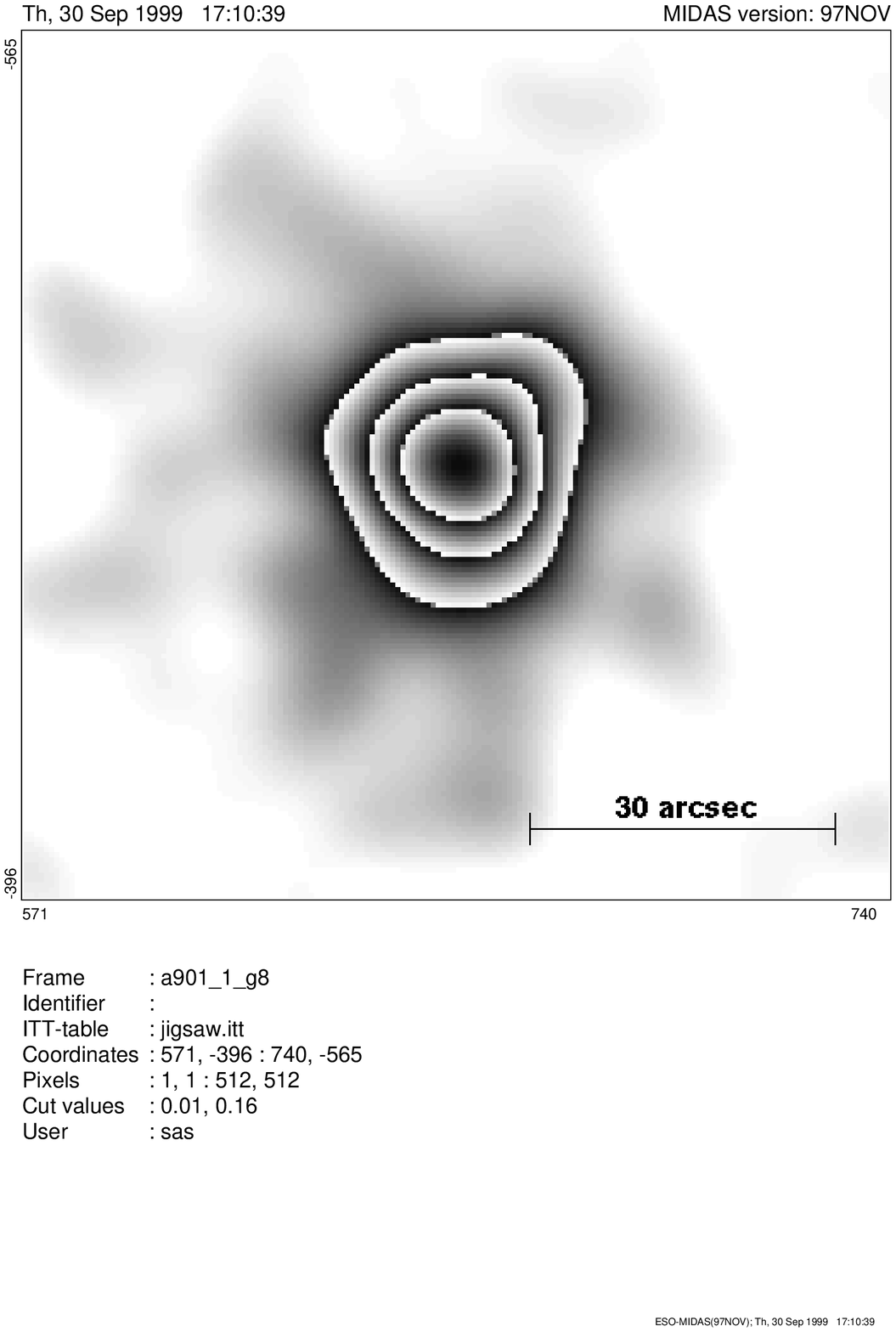,width=8.cm,clip=}
\caption[]{ROSAT/HRI image of A901 smoothed with a Gaussian of
$\sigma=4\arcsec$. This image is a zoom of the dashed square in
Fig.~\ref{fig:a901_over}a. 
The cluster has a very regular and compact
structure. 
}
\label{fig:a901_cen}
\end{figure}

The 
only cluster emission is coming from the region indicated by ``A901''
in Fig.~\ref{fig:a901_over}a 
($\alpha_{2000}=09^{\rm h} 55^{\rm m} 57.0^{\rm s}$, $\delta_{2000}=
-09\degr58\arcmin59\arcsec$).
This emission is shown magnified in
Fig.~\ref{fig:a901_cen} on a scale of 1.4$\arcmin$. The emission is
very compact, but not point-like, as can be clearly seen from the
comparison of the cluster profile and PSF (see
Fig.~\ref{fig:prof}a). A $\beta$-model fit to the cluster profile
(Cavaliere \& Fusco-Femiano 1976; Jones \&
Forman 1984) reveals an extremely
small core radius of $0.10\pm0.03\arcmin$ or $22\pm5$kpc (see also
Table~\ref{tab:summary}) reflecting the compactness of the emission.
%Due to the faintness of the cluster combined with the relatively low
%sensitivity of the ROSAT/HRI, some cluster emission might be
%hidden in the background, so that the true extent might be slightly
%larger. 

The X-ray emission of A901 can be traced out to a radius of almost
2$\arcmin$, 
corresponding to 430 kpc. Within this radius a countrate of
$0.059\pm0.002$ cts/s is found. 
If the emission
of A901 and the 6 point sources  
is summed up, the total countrate is at least a factor 1.8 larger
than the cluster countrate, i.e. the cluster countrate would be 
largely overestimated if the point sources were not resolved.
For the flux and luminosity shown in
Table~\ref{tab:summary} only the cluster emission of A901 was used.

%Using only the cluster emission of A901, the Galactic hydrogen column
%density by Dickey \& Lockman (1990) $n_H = 0.51\times 10^{21}$cm$^2$
%and assuming a temperature of 4 keV from $L_X-T$ relations (Allen \&
%Fabian 1998; Markevitch 1998; Arnaud \&
%Evrard 1999) we find a luminosity and a flux in the ROSAT band of
%$L_X(0.1$-$2.4$keV$)=(3.6\pm0.1)\times 10^{44}$erg/s and
%$f_X(0.1$-$2.4$keV$)=(3.0\pm0.1)\times 10^{-12}$erg/s/cm$^2$,
%respectively. 
%Assuming 3 and 6 keV as lower and upper limits for the temperature,
%respectively, we find a bolometric luminosity
%$L_X(bol)=6.5^{+1.3}_{-0.6}\times 10^{44}$erg/s. 

Estimating a temperature of 4 keV from $L_X-T$ relations (Allen \&
Fabian 1998; Markevitch 1998; Arnaud \&
Evrard 1999) and 
assuming hydrostatic equilibrium we estimate the total mass at
the outer radius $M_{tot}(r<430$kpc$)\approx 9 \times 10^{13} \left ( {T
\over 4{\rm keV} } \right ) \msol$. The gas
mass is $M_{gas}(r<430$kpc$)=1.2 \times 10^{13} \msol$,
i.e. the gas mass fraction is about 13\%.

Obviously, A901 with a flux of
$f_X(0.1$-$2.4$keV$)=3.0\times10^{-12}$erg/s/cm$^2$ is falsely in the
RASS X-ray brightest 
Abell cluster sample of Ebeling et al. (1996) as this sample has a flux
limit of $5\times 10^{-12}$erg/s/cm$^2$. Ebeling et al. list a
flux of $5.2\times 10^{-12}$erg/s/cm$^2$ for the ``brighter subcluster'',
which is in reality not cluster emission. For the ``fainter
subcluster'', which is the true A901 emission, they list correctly
$3.0\times10^{-12}$erg/s/cm$^2$, but 
this value is far below their flux limit.

The compact (but not point-like) nature
of the X-ray emission ($r_c=22$ kpc)
could be interpreted as emission from a galaxy or from a
group of galaxies. But
a comparison of X-ray luminosity and blue
luminosity of the central galaxy (16$^m$ in B) 
shows that A901 lies far above 
the $L_X - L_B$ relation for
early-type galaxies found by Eskridge et al. (1995) and Irwin \&
Sarazin (1998). A group of
galaxies can also be excluded, not only because the X-ray luminosity is
too high, but also from the gas mass fraction.
The gas mass fraction of 13\% is typical for a normal cluster (Ettori \&
Fabian 1999, Schindler 1999), and would be too high for a group
of galaxies (e.g. Pildis et al. 1995). An estimate of the central
cooling time yields about $t_{cool} \approx 
10^9$ years. 
Therefore it is possible, that the compact X-ray emission is 
caused by a cooling flow.

%This
%situation has some similarities to the situation of M49 in the southern
%subcluster of the Virgo cluster. There also the (sub-)cluster emission
%is very compact ($r_c \la 17$ kpc, Schindler et al. 1999) centred
%on the central galaxy which hosts probably a cooling flow (Irwin \&
%Sarazin 1996). 
%But there are also differences: except for the central region, the X-ray
%emission of M49 is very asymmetric -- obviously it is interacting with
%the intra-cluster gas (Irwin \& Sarazin 1996) -- whereas the image of
%A901 is more or less spherically symmetric (see
%Fig.~\ref{fig:a901_cen}). Of course, A901 is further away and
%therefore this fainter emission might be lost in the background.
%The mass of M49 $M_{tot}(r<34 kpc)=1.8 \times 10^{12} \msol$ (Irwin \&
%Sarazin 1996), while the mass of A901 at the same radius 
%$M_{tot}(r<34 $kpc$) \approx 5 \times 10^{12} \left ( {T
%\over 4{\rm keV} } \right ) \msol$ is much larger. Also the X-ray
%luminosity of  A901 is much higher than that of the M49 halo.

\subsection{Abell 1437}

\begin{figure} 
\psfig{figure=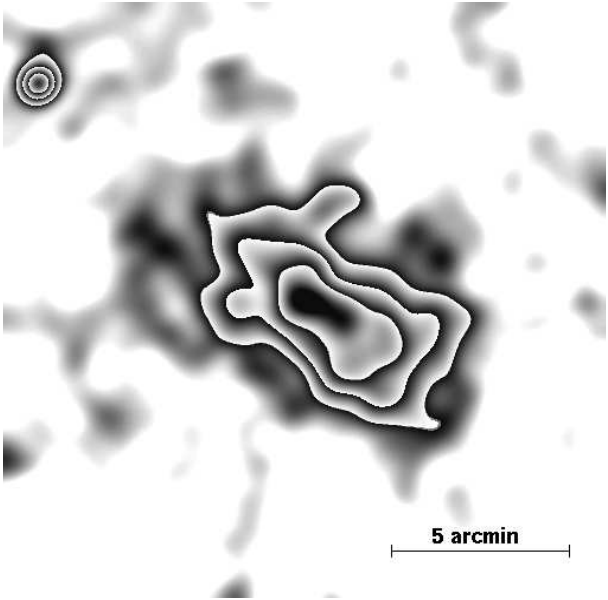,width=8.cm,clip=}
\caption[]{ROSAT/HRI image of A1437 smoothed with a Gaussian of
$\sigma=24\arcsec$. 
The point source in the NE is
probably not connected with the cluster.
}
\label{fig:a1437}
\end{figure}

The cluster A1437 at a redshift z=0.1339 (Struble \& Rood 1987)
is the most X-ray luminous
cluster of this sample.
%Figure~\ref{fig:a1437} shows the X-ray
%emission. 
The cluster centre in X-rays (see Table~\ref{tab:summary})
does not coincide with the optical position: Abell et al. (1989)
determined a position 45$\arcmin$ in the SE of the X-ray maximum.
The emission of the cluster is strongly elongated in SW-NE
direction (see Fig.~\ref{fig:a1437}). 
This elongation can be seen
as well in the RASS. The RASS distinguishes also easily the point
source in the NE, for which an 
optical counterpart can be found on APM finding charts
(see Table~\ref{tab:point}).
Although the cluster shape is not exactly elliptical, we fit
ellipses to the isophotes (Bender \& Moellenhof 1987)
to estimate the elongation. The position angle varies 
around $55^{\circ}$ (N over E). The
minimum axis ratio of 0.38
is reached at 0.01 cts/s/arcmin$^2$. At
this level the centre of the ellipse is shifted 35$\arcsec$ to the
west and 36$\arcsec$ to the south with respect to the position of the
X-ray maximum. 

The fit parameters of the surface brightness profile are not well
constrained (see Table~\ref{tab:summary}) 
because of the
non-spherical morphology of the cluster. Therefore,  
radial profiles of the cluster emission are determined in four
different sectors 
using as centre the X-ray maximum listed in
Table~\ref{tab:fit1437} and subsequently fitted with $\beta$ models
(see Fig.~\ref{fig:prof}b). The two central bins show some
excess emission. This excess cannot come from a cooling flow because the
central cooling time is about $2 \times 10^{10}$ years. It is
probably a small contamination by an AGN. 
Because of this excess we try to fit the overall profile with 
and without these two bins.  The results are listed in 
Table~\ref{tab:fit1437}. In both -- the fit parameters and the fit
curves -- it is
obvious, that the cluster is very asymmetric.

Such asymmetries can arise during a merger of subclusters. From
combined N-body and hydrodynamic simulations it is known that such
elongated morphologies are common shortly after the collision of two
subclusters, when the intra-cluster gas is squeezed out perpendicular
to the collision axis (Schindler \& M\"uller 1993).

\begin{table*}[htbp]
\begin{center}
\begin{tabular}{|c|c|c|c|c|c|}
\hline
region & sector    & inner radius  & $S_0$ & $r_c$ & $\beta$ \\
       & (N over E)& (arcmin [kpc])& ($10^{-2}$ counts/arcmin$^2$/s) 
                                           & (arcmin [kpc])& \\ 
\hline
all    & $  0^\circ - 360^\circ$& 0        &1.2 &2.8 [520] & 0.63\\
all    & $  0^\circ - 360^\circ$& 0.5 [90] &1.2 &3.7 [700] & 0.80\\
NE     & $ 10^\circ - 100^\circ$& 0        &1.2 &2.5 [470] & 0.57\\
NE     & $ 10^\circ - 100^\circ$& 0.5 [90] &1.1 &3.7 [700] & 0.77\\
SE     & $100^\circ - 190^\circ$& 0        &1.5 &0.7 [140] & 0.35\\
SE     & $100^\circ - 190^\circ$& 0.5 [90] &2.4 &0.3 [ 50] & 0.32\\
SW     & $190^\circ - 280^\circ$& 0        &1.4 &$\infty$&$\infty$\\
SW     & $190^\circ - 280^\circ$& 0.5 [90] &1.4 &$\infty$&$\infty$\\
NW     & $280^\circ - 370^\circ$& 0        &1.1 &2.7 [500] & 0.75 \\
NW     & $280^\circ - 370^\circ$& 0.5 [90] &1.1 &2.8 [530] & 0.79 \\
\hline
\end{tabular}
\end{center}
\caption{Fit parameters of the profile of A1437}
\label{tab:fit1437}
\end{table*}

The X-ray emission can be traced out to about 9$\arcmin$. After
excluding the point source in the NE a countrate of $0.25\pm0.01$ cts/s
is found. This corresponds to a flux of
$f_X(0.1$-$2.4$keV$)=(1.04\pm0.03)\times 10^{-11}$erg/s/cm$^2$.
%Assuming a temperature of $T=6^{+3}_{-2}$ keV from $L_X-T$
%relations and a Galactic
%hydrogen column density of $n_H = 0.187 \times 10^{21}$ cm$^{-2}$ (Dickey \&
%Lockman  1990) this countrate corresponds to a luminosity 
%$L_X(0.1$-$2.4$keV$)=(8.5\pm0.3)\times 10^{44}$erg/s and a flux 
%$f_X(0.1$-$2.4$keV$)=(1.04\pm0.03)\times 10^{-11}$erg/s/cm$^2$ in the ROSAT
%band. The estimated bolometric luminosity is 
%$L_X(bol)=19^{+5}_{-3}\times 10^{45}$erg/s. 
For A1437, which is the most luminous cluster of this sample, the flux
determination 
from the RASS ($f_X(0.1$-$2.4$keV$)=1.02\times 10^{-11}$erg/s/cm$^2$
(Ebeling et al. 1996) and 
$f_X(0.1$-$2.4$keV$)=1.00\times 10^{-11}$erg/s/cm$^2$ (Ebeling
et al. 1998), respectively) is reliable. Also the morphological
determination from the RASS is good: the point source in the NE can be
distinguished easily and the elongated shape of the cluster is visible
in the RASS as well.

\subsection{Abell 3570}

\begin{figure*} 
\begin{tabular}{cc}   
\psfig{figure=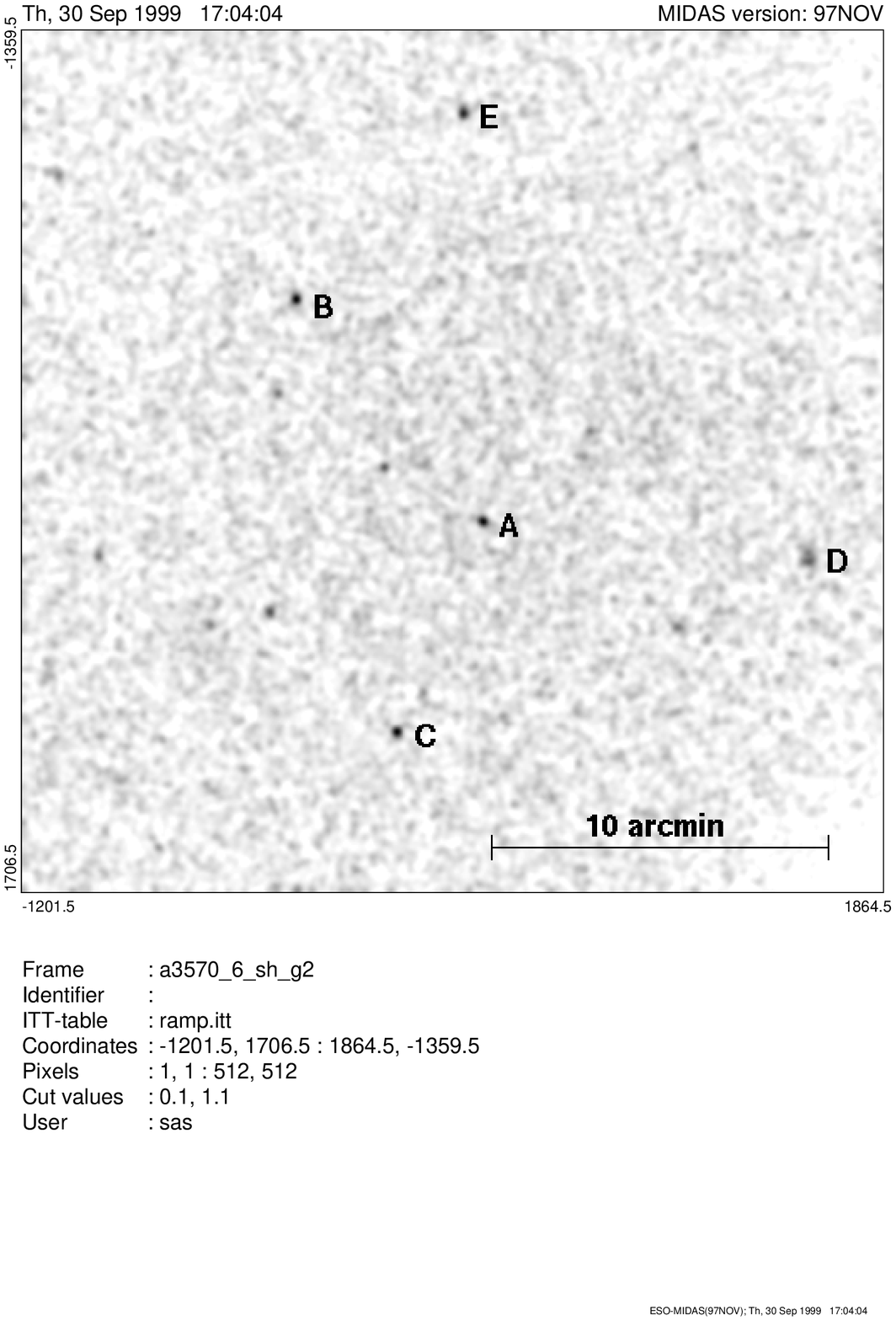,width=8.cm,clip=} \put(-210.,200.){a} &
\psfig{figure=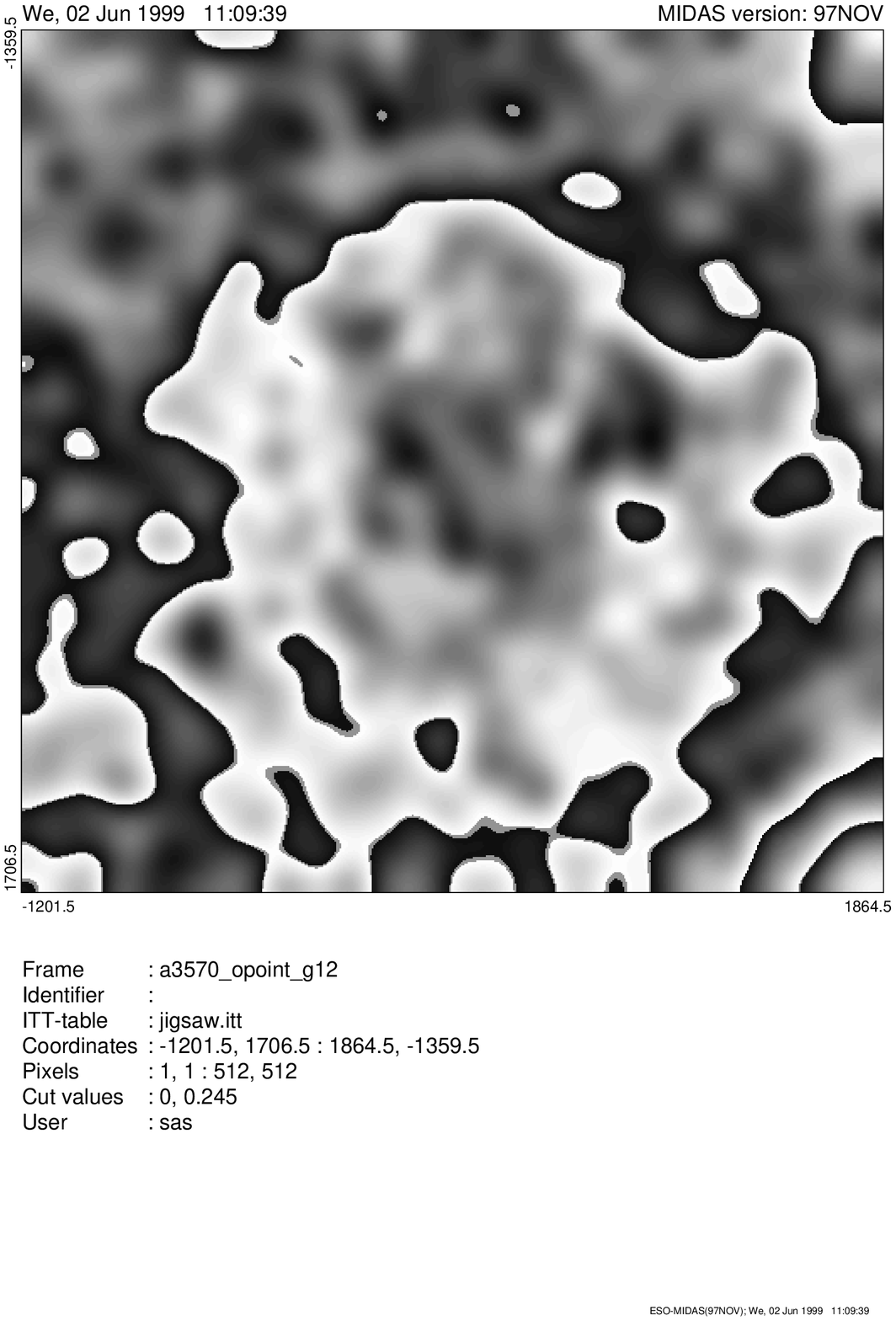,width=8.cm,clip=} \put(-210.,200.){b} 
\cr
\end{tabular}
\caption[]{
ROSAT/HRI image of A3570. a) In an image smoothed with a Gaussian of
$\sigma=6\arcsec$ distinct sources dominate over the faint cluster
emission. The sources A, B, C, and E are point-like while D has a
small extent. These sources are listed in
Table~\ref{tab:point}. 
b) After subtraction of the five point sources a strongly smoothed image
($\sigma=36\arcsec$) shows the cluster emission. The scale is the same
as in a). At the upper and lower right corner the edge of the
field-of-view is visible.
}
\label{fig:a3570}
\end{figure*}

The cluster A3570 is the 
nearest cluster of this sample ($z=0.037$). 
%It is a richness class 0
%cluster. 
The X-ray emission is faint and the extent is of the same
order as the field-of-view of the ROSAT/HRI. With small
smoothing the cluster X-ray emission is hardly visible, because the
region is
dominated by discrete sources (see Fig.~\ref{fig:a3570}a and
Table~\ref{tab:point}). One of the sources (D) is not point-like but
has a small extent. This source can be identified with the galaxy
ESO 325 - G016 -- a cluster galaxy at redshift of $z=0.03795$
(Postman \& Lauer 1995). 
%For the other sources no optical counterparts
%can be found.
% on the ROE/NRL COSMOS finding charts.
 
To make the cluster emission visible we remove all point sources,
which have a signal-to-noise ratio of at least 3 
above the surrounding cluster emission,
by fitting a warped
surface to the pixels surrounding the point source region and apply
a much coarser smoothing (see Fig.~\ref{fig:a3570}b). 
The cluster emission is extended and
regular. There is no significant sign of subclustering or merging, i.e. the
complex structure seen in the RASS disappears on resolving the
discrete sources. Therefore, A3570 is very likely a relaxed cluster.
Fitting the
profile for this cluster is not possible because the profile is so
shallow. 

\begin{table*}[htbp]
\begin{center}
\begin{tabular}{|c|c|c|c|c|c|c|c|c|}
\hline
pointing & source & $\alpha$(J2000) & $\delta$(J2000) &HRI count rate& identification \\
\hline
A901    & A & 09 56 28.2      & -09 57 15& 0.039& galaxy 17$^m$(blue) \\
        & B & 09 56 35.4      & -10 04 53& 0.006& -       \\ 
        & C & 09 56 30.6      & -10 00 12& 0.001& star 14$^m$(blue)\\
        & D & 09 56 18.1      & -09 53 57& 0.002& -       \\ 
        & E & 09 56 22.3      & -09 55 07& 0.002& star 15$^m$(blue) or galaxy 17$^m$(blue) \\
        & F & 09 56 35.6      & -10 10 08& 0.001& -       \\ 
\hline
A1437   & A & 12 00 55.7      & 03 26 58& 0.007 & galaxy 16$^m$(red)\\
\hline
A3570   & A & 13 47 12.5      & -37 57 15& 0.002& -        \\
        & B & 13 47 40.7      & -37 50 38& 0.002& -        \\
        & C & 13 47 25.6      & -38 03 30& 0.002& -        \\
        & D & 13 46 23.3      & -37 58 21& 0.001& ESO 325 - G016 \\
        & E & 13 47 15.5      & -37 45 06& 0.002& -        \\      
\hline
\end{tabular}
\end{center}
\caption{Positions and identification of non-cluster X-ray 
sources in the pointings}
\label{tab:point}
\end{table*}

Because of the large extent of the cluster the image had to be
vignetting corrected for the countrate determination (Snowdon 1998). 
The countrate
determination is difficult, because the cluster emission
fills probably the whole field-of-view of the HRI.
We estimate the countrate to
be $1.0^{+0.4}_{-0.7}$ cts/s by counting all the photons within
$r=15\arcmin$ (corresponding to 250 kpc), 
excluding the discrete sources and using a standard
ROSAT/HRI background. Out to this radius we can
clearly trace the X-ray emission, but probably the cluster extends
further beyond the field-of-view. Therefore, we estimate the upper
limit of the countrate by adding the photons found beyond this radius
north and east of the cluster and assume the same number in the
south and west, which is not covered by the detector.
The lower limit is obtained by using
the background at the border of the field-of-view, which is a very
conservative estimate. For flux and luminosity see
Table~\ref{tab:summary}.  

The discrete sources change the cluster morphology
drastically by feigning substructure in the RASS image.
But they do not
contribute significantly to the countrate. We estimate the countrate
of the discrete sources 
by fitting a warped surface to the pixels surrounding the point source region
and subtract these fitted counts from the original counts.
The discrete sources together have a very small count rate of about
0.01 cts/s,
which is negligible compared to the cluster emission.

%Assuming a temperature of $T=4$ keV from $L_X-T$
%relations and a Galactic
%hydrogen column density of $n_H = 0.44 \times 10^{21}$ cm$^{-2}$ (Dickey \&
%Lockman  1990) the cluster countrate corresponds to a luminosity 
%$L_X(0.1$-$2.4$keV$)=(3.2^{+1.1}_{-2.3})\times 10^{44}$erg/s and a flux 
%$f_X(0.1$-$2.4$keV$)=(5.2^{+1.8}_{-3.8})\times 10^{-12}$erg/s/cm$^2$ 
%in the ROSAT
%band. Assuming an error range of 3-6 keV for the temperature we
%estimate the bolometric luminosity 
%$L_X(bol)=5.7^{+3.8}_{-4.3}\times 10^{44}$erg/s. 

For the X-ray luminosity 
$L_X(0.1$-$2.4$keV$)=(3.2^{+1.1}_{-2.3})\times 10^{44}$erg/s the 
velocity dispersion $\sigma=46$0 km/s is relatively 
low. While a temperature of 4 keV is consistent with the $L_X-T$
relations, the $\sigma-T$ relations predict only 2
keV (White et al. 1997; Mushotzky \& Scharf 1997, Wu et al. 1999).
The small velocity dispersion confirms the conclusion from the ROSAT/HRI
observation, that A3570 is a regular, non-merger cluster.

\begin{table*}[htbp]
\begin{center}
\begin{tabular}{|c|c|c|c|c|}
\hline
cluster &                               & A901   & A1437  & A3570  \\
\hline
\smash{\lower6pt\hbox{position (J2000)}} 
                              & $\alpha$& 09 55 57.0      
                                        & 12 00 25.7 
                                        & 13 47 16.1 \\
                              & $\delta$& -09 58 59       
                                        & +03 20 50  
                                        & -37 56 28  \\
%\multispan2
                HRI countrate &  [cts/s]& $0.059\pm0.002$ 
                                        & $0.25\pm0.01$
                                        & $1.0^{+0.4}_{-0.7}$ \\
%\multispan2
luminosity (0.1-2.4 keV)&[$10^{44}$ erg/s]&$3.6\pm0.1$      
                                        & $8.5\pm0.3$
                                        & $3.2^{+1.1}_{-2.3}$ \\
%\multispan2
flux (0.1-2.4 keV)&[$10^{-12}$ erg/cm$^2$/s]&$3.0\pm0.1$      
                                        & $10.4\pm0.3$
                                        & $5.2^{+1.8}_{-3.8}$ \\
%\multispan2
luminosity (bolometric)&[$10^{44}$ erg/s]&$6.5^{+1.3}_{-0.6}$
                                        & $19^{+5}_{-3}$
                                        & $5.7^{+3.8}_{-4.3}$ \\ 
%\multispan2
$S_0$&[cts/s/arcmin$^2$]&0.27          & 0.012                 & - \\
%\multispan2
$\beta$               && $0.50\pm0.03$ & $0.80^{+0.75}_{-0.25}$& - \\
%\multispan2
$r_c$&[arcmin]         & $0.10\pm0.03$ & $3.8^{+2.5}_{-1.4}$   & - \\
%\multispan2 
$r_c$&[kpc]            & $22\pm5$    & $710^{+470}_{-260}$   & - \\
%\multispan2
redshift              && 0.17          & 0.13                  & 0.037\\
velocity dispersion   & [km/s]&   -    &  -                    & 460  \\      
%\multispan2
$n_H$&[$10^{21}$cm$^{-2}$]&0.51        & 0.19                  & 0.44 \\
\hline
\end{tabular}
\end{center}
\caption{Summary of the properties of the three clusters. For the
flux and luminosity calculation hydrogen
column densities from Dickey \& Lockman (1990, see last line) were used and 
temperatures were estimated from $L_X-T$ relations (Allen \&
Fabian 1998; Markevitch 1998; Arnaud \&
Evrard 1999): $T=4^{+2}_{-1}$keV for A901 and A3570,
$T=6^{+3}_{-2}$keV for A1437.
}
\label{tab:summary}
\end{table*}

\section{Conclusions}

We analysed three clusters of galaxies with low to intermediate 
X-ray luminosities which
show an irregular appearance  in the ROSAT All-Sky Survey. 
As the confusion with fore- and background sources is
increasingly critical with decreasing flux and increasing
substructure of the cluster, we test the limitations of the RASS.
With only three clusters, of course,
this cannot be a statistical study, but it is meant to be
a ``worst-case'' study. 
We would like to stress that for higher fluxes source confusion is
less important and that for clusters with these fluxes the RASS
results are very reliable.

The results of the follow-up observations -- X-ray observations with the 
ROSAT/HRI and optical spectroscopic observations -- are summarized
in Table~\ref{tab:summary}. 
The ROSAT/HRI observations, which have a much better spatial
resolution than the RASS, revealed in two of the three cases a
different morphology than seen in the RASS. The reasons are point
sources which could not be resolved in RASS and were therefore
confused with the cluster emission. In one of the three clusters the
countrate measurement is affected by the point sources, in the two
other cases the countrate measurements of the RASS are reliable.
The results for the three clusters are the following:
\begin{itemize}
\item Abell 901: With a flux of $f_X(0.1$-$2.4$keV$)=3.0\pm0.1 \times
10^{-12}$erg/cm$^2$/s this cluster is the faintest one in the
sample. Both the measurement of the countrate and the determination of
the morphology provide difficulties in the RASS. Several discrete
sources cannot be distinguished from the cluster emission in the
RASS. With the RASS information only it was concluded that
this cluster consists of two subclumps (Ebeling et al. 1996), 
while the true cluster
X-ray emission is very compact ($r_c=22$kpc). The countrate would be 
overestimated by at least 80\% if the discrete sources cannot be
separated from the cluster. 

\item Abell 1437: This cluster has the highest flux of all the clusters in
the sample $f_X(0.1$-$2.4$keV$)=1.04\pm0.03 \times 10^{-11}$erg/cm$^2$/s. 
With such a high flux the RASS results are very
reliable. There is
no problem with the countrate or morphology determination. The point source
in the NE of the cluster can  easily be distinguished in the RASS.

\item Abell 3570: This cluster has several
point sources superposed on the cluster emission. Although the
countrate determination is not affected by these sources (their 
countrate is only 1\% of the cluster countrate) the true morphology
is very regular in contrast to the impression from the
RASS. Optical measurements of the cluster galaxies give a very small
velocity dispersion confirming the picture of a relaxed cluster.
\end{itemize}
We conclude that for clusters with fluxes less than a few times
$10^{-12}$erg/cm$^2$/s, which have at the same time an irregular
morphology, the confusion with fore- and background sources can be a
problem in the RASS. So morphological analyses of RASS clusters tend
to overestimate the fraction of clusters with substructure. 

A new X-ray all-sky survey with a potential 
second ABRIXAS (Friedrich et al. 1996) 
mission would have several advantages to
push this source confusion limit down.
\begin{itemize}
\item The spatial resolution of ABRIXAS 
is about 2.5 times better than the resolution of the
RASS, which makes it easier to distinguish between point and extended
sources. 

\item The energy range of ABRIXAS (0.5-12 keV) is much better suited for
clusters than the ROSAT range (0.1-2.4 keV). Although both surveys
have about the same sensitivity at 1 keV, ABRIXAS with its harder
energy range would detect  3-5 times more photons from a standard
cluster. 

\item The spectral resolution of ABRIXAS is better than that of the
ROSAT/PSPC. Together with the wider energy range of ABRIXAS this
provides an improved possibility to separate hard cluster emission
from soft foreground sources.
\end{itemize}

\begin{acknowledgements}

I thank Chris Collins for introducing me into the secrets of optical
spectra, Hans B\"ohringer for making the RASS images of the three
clusters available to me, Peter Friedrich for providing the ABRIXAS
numbers, and Phil James and Joachim Wambsganss for carefully reading
the manuscript.  
It is a pleasure to thank 
Carlo Izzo for his most helpful EXSAS support.
I acknowledge gratefully the hospitality of the Institut d'Estudis
Espacials de Catalunya in Barcelona. During 
the stay there I was supported by the TMR grant ERB-FMGE CT95 0062 by
CESCA-CEPBA. 
\end{acknowledgements}
%
%________________________________________________________________
%


\begin{thebibliography}{}
\bibitem{} Abell G.O., Corwin H.G., Olowin R.P., 1989, ApJS 70, 1
\bibitem{} Allen S.W., Fabian A.C., 1998, MNRAS 297, L57
\bibitem{} Arnaud M., Evrard A.E., 1999, MNRAS 305, 631
\bibitem{} Bender R., Moellenhof C., 1987, A\&A 177, 71 
\bibitem{} Cavaliere A., Fusco-Femiano R., 1976, A\&A 49, 137
%\bibitem{} De Grandi S., Guzzo L., B\"ohringer H., Molendi S., Chincarini
%     G., Collins C., Cruddace R., Neumann D., Schindler S., Schuecker
%     P., Voges W., 1999, ApJ 513, L17
\bibitem{} Dickey J.M., Lockman F.J., 1990, ARA\&A 28, 215  
\bibitem{} Ebeling H., Voges W., B\"ohringer H., Edge A.C., Huchra
     J.P., Briel U.G., 1996, MNRAS 281, 799 
\bibitem{} Ebeling H., Edge A.C., B\"ohringer H., Allen S.W., Crawford
     C.S., Fabian A.C., Voges W., Huchra J.P., 1998, MNRAS 301, 881 
\bibitem{} Eskridge P.B., Fabbiano G., Kim D.-W., 1995, ApJS 97, 141
\bibitem{} Ettori S., Fabian A.C., 1999, MNRAS 305, 834
\bibitem{} Felenbok P., G\'uerin J., Fernandez A., Cayatte V.,
    Balkowski C., Kraan-Korteweg R.C., 1997, Experimental Astronomy  
    7, 65 
\bibitem{} Friedrich P., Hasinger G., Richter G., et al.,
%Fritze K., Tr\"umper
%J., Br\"auninger H., Predehl P., Staubert R., Kendziorra E., 
    1996, in:
    proceedings of the conference "R\"ontgenstrahlung from 
    the universe", MPE Report 263, H.U. Zimmermann, J.E. Tr\"umper,
    H. Yorke (eds.), p. 681 
\bibitem{} Fukazawa Y., Makishima K., Tamura T., Ezawa H., Xu H.,
        Ikebe Y., Kikuchi K., Ohashi T., 1998, PASJ 50, 187
\bibitem{} Girardi M., Giuricin G., Mardirossian F., Mezzetti M.,
     Boschin W., 1998, ApJ 505, 74 
%\bibitem{} Irwin J.A., Sarazin C.L., 1996, ApJ 471, 683
\bibitem{} Irwin J.A., Sarazin C.L., 1998, ApJ 499, 650
\bibitem{} Jones C., Forman W., 1984, AJ 276, 38
\bibitem{} Pildis R.A., Bregman J.N., Evrard A.E., 1995, ApJ 443, 514
\bibitem{} Ponman T.J., Bourner P.D.J., Ebeling H., B\"ohringer H.,
      1996, MNRAS 283, 690 
\bibitem{} Postman M., Lauer T.R., 1995, ApJ 440, 28
\bibitem{} Markevitch M., 1998, ApJ 504, 27
\bibitem{} Mohr J.J., Mathiesen B., Evrard A.E., 1999, ApJ 517, 627 
\bibitem{} Mushotzky R.F., Scharf C.A., 1997, ApJ 482, L13 
\bibitem{} Reiprich T.H., B\"ohringer H., 1999, in: Proceedings of
        the 4$^{th}$ ASCA Symposium on Heating and Acceleration in the 
        Universe held in Tokyo, Japan, Inoue H., Ohashi T., Takahashi
        T. (eds.), in press
\bibitem{} Schindler S., 1999, A\&A 349, 435
\bibitem{} Schindler S., M\"uller E., 1993, A\&A 272, 137
\bibitem{} Schindler S., Wambsganss J., 1996, A\&A 313, 113 
\bibitem{} Schindler S., Belloni P., Ikebe Y., Hattori M., Wambsganss J.,
       Tanaka Y., 1998, A\&A, 338, 843
%\bibitem{} Schindler S., Binggeli B., B\"ohringer H., 1999, 
%    A\&A 343, 420
\bibitem{} Snowden S.L., 1998, ApJS 117, 233
\bibitem{} Struble M.F., Rood H.J., 1987, ApJS 63, 543
\bibitem{} Tr\"umper J., 1983, Adv. Space Res. 2, 142
\bibitem{}  Voges W., Boller T., Dennerl K., et al., 
%Englhauser J., Gruber R.,
%Haberl F., Paul J., Pietsch W., Tr\"umper J.E., Zimmermann H.U., 
        1996,
        in: Proceedings of the workshop R\"ontgenstrahlung from the
        universe, held in W\"urzburg, Germany, Zimmermann H.U.,
        Tr\"umper J.E., Yorke H. (eds.), MPE Report 263, p. 637
\bibitem{} White D.A., Jones C., Forman W., 1997, MNRAS 292, 419
\bibitem{} Wu X.-P., Xue Y.-J., Fang L.-Z., 1999, astro-ph/9905106

\end{thebibliography}
\end{document}